\documentclass{article}
\usepackage[english]{babel}
\usepackage{amsmath}
\usepackage{amssymb}
\usepackage[utf8]{inputenc}
\usepackage{graphicx}
\usepackage{tikz}
\usepackage{cite}
\graphicspath{{pictures/}}

\numberwithin{equation}{section}

\title{Parametric Instabilities in Shallow Water Magnetohydrodynamics Of Astrophysical Plasma in External Magnetic Field}
\date{}
\author{
    Klimachkov D.A.$^{1}$,Petrosyan A.S.$^{1,2}$\\
    $^{1}$Space Research Institute of Russian Academy of Science,\\
    84/32, Profsoyuznaya str., Moscow, Russia, 117997\\
    $^{2}$Moscow Institute of Physics and Technology (State University),\\
    9 Institutskyi per., Dolgoprydny, Moscow Region, Russia, 141700
}

\begin{document}
\maketitle

\begin{abstract}
This letter discusses rotating magnetohydrodynamics (MHD) of a thin layer of astrophysical plasma. To describe a thin plasma layer with a free surface in a vertical external magnetic field we use the shallow water approximation. The presence of a vertical magnetic field essentially changed the wave processes dynamics in astrophysical plasma compared to the neutral fluid and plasma layer in a thoroidal magnetic field. In present case thre are three-waves nonlinear interactions. Using the asymptotic multiscale we deduced nonlinear wave packets interaction equations: three magneto-Poincare waves interaction, three magnetostrophic waves interaction, the interaction of two magneto-Poincare and one magnetostrophic wave and two magnetostrophic and one magneto-Poincare wave interaction. The existence of decay instabilities and parametric amplifications is predicted. We found following four types of decay instabilities: magneto-Poincare wave decays into two magneto-Poincare waves, magnetostrophic wave decays into two magnetostrophic waves, magneto-Poincare wave decays into one magneto-Poincare wave and one magnetostrophic wave, magnetostrophic wave decays into one magnetostrophic wave and one magneto-Poincare wave. Also following mechanisms of parametric amplifications were found: parametric amplification of magneto-Poincare waves, parametric amplification of magnetostrophic waves, magneto-Poincare wave amplification in magnetostrophic wave presence nd magnetostrophic wave amplification in magneto-Poincare wave presence. The instabilities growth rates and parametrical amplifications factors were found respectively.
\end{abstract}

\section{Introduction}
Plasma in various stars and planets is described by MHD of thin fluid layer with a free surface in the gravity field. For example, solar tachocline flows (a thin layer inside the sun is above the convective zone) \cite{zaqarashvili2007rossby}, neutron stars atmosphere dynamics \cite{heng2009magnetohydrodynamic}, accreting matter flows in the neutron stars \cite{inogamov2010spread}, tidally synchronized exoplanets with magnetoacctive atmospheres \cite{hengannualreview}, \cite{hengexoplanetatmospheres}, \cite{cho2008atmospheric}. To describe such flows of astrophysical plasma MHD shallow water approximation \cite{gilman1967stability} and quasigeostrophic MHD approximation \cite{gilman1967stability}, \cite{tobias2007beta}, \cite{balk} are used. The present letter is devoted to the study of weakly nonlinear waves interactions in shallow water magnetohydrodynamics. We consider a thin fluid layer (comparing to the characteristic horizontal linear scale) and neglect vertical accelerations. In this case rotating MHD equations in shallow water approximation  are the alternative for heavy fluid MHD equations with a free surface.  We consider a thin layer of incompressible inviscid fluid with a free surface in the gravity field. The frame of reference is non-inertial and rotates with fluid layer. MHD shallow water equations are obtained from general incompressible MHD equations by depth-averaging. The pressure is assumed to be hydrostatic and layer height is assumed to be much smaller than characteristic horizontal scale \cite{gilman1967stability}, \cite{gilman2000magnetohydrodynamic}, \cite{Tarasevich}, \cite{karelsky2013nonlinear}, \cite{karelsky2014english}, \cite{de2001hyperbolic}, \cite{dellar2003dispersive}, \cite{zeitlin2013remarks}. The obtained system plays important role in astrophysical plasma studies same as classical shallow water equations in neutral flows. In linear approximation MHD shallow water equations have the solutions describing magnetogravity (or magneto-Poincare) waves and magnetostrophic waves. In neutral hydrodynamics in shallow water approximation there are only garvitational Poincare waves. The Poincare waves dispersion relation and nonlinear terms  in neutral fluid shallow water equations exclude Poincare waves interactions in case of weakly nonlinear finite amplitude waves \cite{vallis2006atmospheric}, because in this case phase matching conditions are not fulfilled. In solar tachocline flows thoroidal magnetic fiel takes place. The presence of thoroidal magnetic field leads to two types of waves: fast mgnetogravity waves (similar to Poincare waves in neutral fluid) called magneto-Poincare and slow alfven waves. In flows with thoroidal magnetic field the dispersion relation doesn't provide phase matching conditions in weakly nonlinear interactions \cite{zaqarashvili2007rossby},\cite{gilman1967stability}, \cite{gilman2000magnetohydrodynamic}. In this study we consider MHD shallow water equations in the external vertical magnetic field. Such configuration of magnetic field is typical for the neutron stars \cite{heng2009magnetohydrodynamic}, \cite{cho2008atmospheric} and for the exoplanets \cite{cho2008atmospheric}. In this case in MHD shallow water approximation new additional terms appear. In linear approximation they describe two types of fast waves: magneto-Poincare waves and magnetostrophic waves \cite{heng2009magnetohydrodynamic}. The present letter is the generalization of the linear theory of MHD shallow water flows developed in \cite{heng2009magnetohydrodynamic} to the case of finite amplitude waves in weakly nonlinear approximation. It is shown that the dispersion relations of linear waves in external vertical magnetic field provide the phase matching conditions needed for nonlinear interactions. In the absence of the external vertical magnetic field magnetostrophic waves disappear and only magneto-Poincare waves remain in neutral fluid hydrodynamics in the gravity field with a free surface. We investigate the interactions of wave packets in rotating shallow water MHD. The analysis of the dispersion relations for both modes shows that there are several types of three waves interactions: three magneto-Poincare waves, three magnetostrophic waves and also intermode interactions: two magneto-Poincare waves and magnetostrophic wave, two magnetostrophic waves and magneto-Poincare wave. To describe nonlinear interactions  we use multiscale asymptotic method \cite{ostrovsky2014asymptotic}. For each mentioned case nonlinear equations of wave amplitudes interaction are derived. The analysis of obtained nonlinear equations which describe three-waves interactions shows that there are two types of instabilities: decay instabilities and parametric wave amplification. We found following four types of decay instabilities: magneto-Poincare wave decays into two magneto-Poincare waves, magnetostrophic wave decays into two magnetostrophic waves, magneto-Poincare wave decays into one magneto-Poincare wave and one magnetostrophic wave, magnetostrophic wave decays into one magnetostrophic wave and one magneto-Poincare wave. The instability growth rates were found. Also following four types of parametrical amplification mechanisms were investigated: parametric amplification of magneto-Poincare waves, parametric amplification of magnetostrophic waves, magneto-Poincare wave amplification in magnetostrophic wave presence and magnetostrophic wave amplification in magneto-Poincare wave presence. Increments are found for each type of instability.

In section 2 we provide MHD shallow water equations on a flat surface with rotation. Linear solutions and dispersion curves are analyzed and conclusions about phase matching presense providing three-waves interaction possibilities in finite amplitude approximaton are made.

In section 3 we deduce the equations for the slowly varying amplitudes of three-waves interactions in shallow water MHD in the external vertical magnetic field.

In section 4 the obtained equations of three-waves interactions are used to analyze physical effects of weakly nonlinear interactions of magneto-Poincare and magnetostrophic waves. Decay instabilities and parametric amplifications are analyzed.

In section 5 the obtained results are formulated.

\section{Qualitative analysis of MHD flows of astrophysical plasma in shallow water approximation}

The MHD shallow water equations on the flat surface are following \cite{heng2009magnetohydrodynamic}:
\begin{equation}\label{6}
\frac{\partial  h}{\partial t}+\frac{\partial  h v_x}{\partial x}+\frac{\partial  h v_y}{\partial y}=0
\end{equation}
\begin{equation}\label{7}
\frac{\partial ( h v_x)}{\partial t}+\frac{\partial ( h (v_x^2-B_x^2))}{\partial x}+g  h \frac{\partial h}{\partial x}+\frac{\partial (\ h (v_x v_y-B_x B_y))}{\partial y}+B_0B_x=2 \omega  h v_y
\end{equation}
\begin{equation}\label{8}
\frac{\partial ( h v_y)}{\partial t}+\frac{\partial ( h (v_x v_y-B_x B_y))}{\partial x}+\frac{\partial ( h ( v_y^2-B_y^2))}{\partial y}+g  h \frac{\partial h}{\partial y}+B_0B_y=-2 \omega  h v_x
\end{equation}
\begin{equation}\label{9}
\frac{\partial ( h B_x)}{\partial t}+\frac{\partial (h (B_x v_y-B_y v_x))}{\partial y}+B_0v_x=0
\end{equation}
\begin{equation}\label{10}
\frac{\partial ( h B_y)}{\partial t}+\frac{\partial ( h (B_y v_x-B_x v_y))}{\partial x}+B_0v_y=0
\end{equation}

In (\ref{6}-\ref{10}) $h$ is  free surface vertical coordinate, $v_x, v_y$ are horizontal velocities in shallow water approximation in the  $xy$-plane, $B_x, B_y$ are horizontal components of depth-averaging magnetic fields in shallow water approximation in the $x$ and $y$ directions respectively, $B_0$ is an external magnetic field which is directed perpendicular to the $xy$-plane, $\omega$ is an angular velocity of plasma. The first equation of the system (\ref{6}-\ref{10}) is continuity equation, the second and the third ones are the equations of conservation of momentum, the fourth and the fifth ones are the equations of magnetic field variations.

MHD shallow water equations on a flat surface are obtained from complete MHD equations system. The thin fluid layer with a free surface is considering in the gravity field in the coordinate system where z-axis is parallel to the gravity vector and has the opposite direction. We assume the fluid layer to be thin and the complete pressure (hydrodynamic and magnetic) to be hydrostatic. The system is integrated by the vertical axis. The presence of external magnetic field $B_0$ changes the boundary conditions are used in deriving (\ref{6}-\ref{10}) \cite{heng2009magnetohydrodynamic}. After averaging we neglect the squares of the deviation values of the velocity and magnetic field from the depth-averaging values \cite{karelsky2013nonlinear}, \cite{Tarasevich}, \cite{karelsky2014english}, \cite{zeitlin2013remarks}.
The dispersion equation of the linear waves in (\ref{6}-\ref{10}) is following:
\begin{equation}\label{22}
\omega^5-\omega^3 (ghk^2 + f^2+2(\frac{B_0}{h})^2)+(\frac{B_0}{h})^2(ghk^2+(\frac{B_0}{h})^2)\omega=0
\end{equation}
In this case because of vertical magnetic field presence there are two types of waves. The first type is the generalization of linear Poicare waves in classical shallow water equations for neutral fluid and has following dispersion relation:
\begin{equation}
\omega_{1,3}=\pm \sqrt{\frac{ghk^2}{2}+\frac{f^2}{2}+(\frac{B_0}{h})^2 + \frac12 \sqrt{ghk^2(ghk^2+2f^2)+f^2(f^2+4(\frac{B_0}{h})^2)}}
\end{equation}
where sign $+$ is for the wave that propagate along vector $\mathbf{k}$, and sign $-$ is for the wave that propagate in opposite direction. The obtained linear solutions are called magneto-Poincare waves. The second type of the linear solutions describes magnetostrophic waves that do not have the analoges in neutral fluid. It has following dispersion relation:
\begin{equation}
\omega_{2,4}=\pm \sqrt{\frac{ghk^2}{2}+\frac{f^2}{2}+(\frac{B_0}{h})^2 - \frac12 \sqrt{ghk^2(ghk^2+2f^2)+f^2(f^2+4(\frac{B_0}{h})^2)}}
\end{equation}
where sign $+$ is for the wave that propagate along $\mathbf{k}$, and sign $-$ is for the wave that propagate in the opposite direction. Subscripts $1,2$ correspond to the waves propagating along $\mathbf{k}$,and subscripts $3,4$ correspond to the waves propagating in the opposite direction for the convinience of considering nonlinear interactions.
The general solution is the sum of the four linear waves.

\begin{figure}
\center{\includegraphics{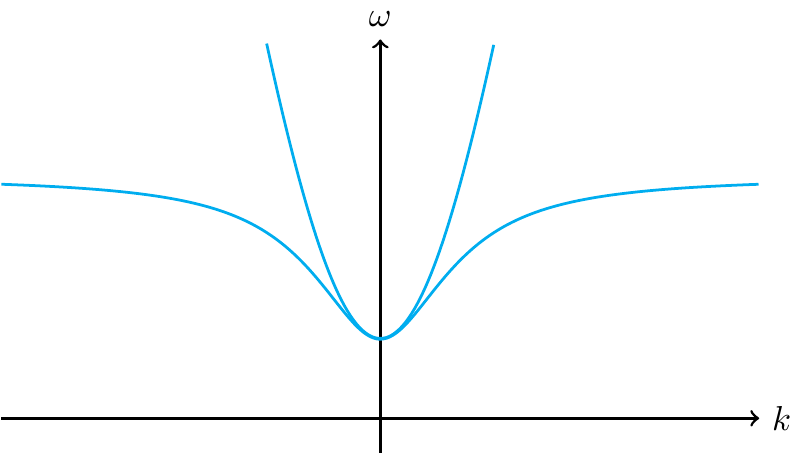}}
\caption{Dispersion curves. The upper one is Poincare mode, the bottom one is magnetostrophic mode}
\label{pic.disperslines}
\end{figure}

The general form of the dispersion curves in case when $\omega>0$ is shown at figure \ref{pic.disperslines}. The upper curve displays magneto-Poincare mode, the bottom curve displays magnetostrophic mode. On the vertical axis is the frequency $\omega$, on the horizontal axis is $k_x$, when the wave vector $\mathbf{k}$ is directed along the $x$-axis. Dispersion relation $\omega(\mathbf{k})$ defines the dispersion surface $\omega(k_x,k_y)$. This surface is cylindrically symmetric because the dispersion relation is symmetric with respect to $k_x$ and $k_y$.
For $\omega<0$ (solutions $\omega_3$ and $\omega_4$) dispersion surface is symmetric with respest to the $\omega=0$ plane.
To assess the possibility of interwaves interactions of described types of waves we need to analyze its dispersion relations and determine the dispersion curves asymptotics for both waves types. The phase matching condition nessesary for interactions of waves with different wave vectors and different frequencies in general case is as follows: $\omega(\mathbf{k_1})+\omega(\mathbf{k_2})=\omega(\mathbf{k_1}+\mathbf{k_2})$. We represent this condition on the graph. The first term defines a point $(k_1,\omega(k_1))$ which is on the dispersion curve of one solution. The second term defines a point $(k_2,\omega(k_2))$ which is on the dispersion curve of another solution. If the second dispersion curve shifted by $(k_1,\omega(k_1))$crosses the first one in a point $(k_3,\omega(k_3))$ it causes the existence of such wave vector $k_2$ that the phase matching condition is satisfied.

\begin{figure}
\center{\includegraphics{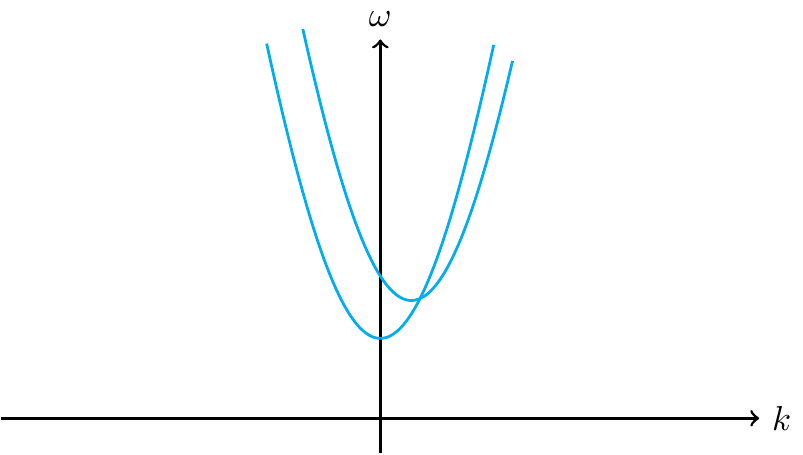}}
\caption{The intersection of the dispersion curves of magneto-Poincare mode}
\label{pic.threemagnetopoincare}
\end{figure}

For the magneto-Poincare waves the phase matching condition is as follows: $\omega_1(\mathbf{k_1})$ $+\omega_1(\mathbf{k_2})=\omega_1(\mathbf{k_1}+\mathbf{k_2})$ where the subscript $1$ corresponds forward magneto-Poincare wave. At low frequencies the dispersion surfaces for the magneto-Poincare mode on the $k_x,k_y$-plane are convex. Figure \ref{pic.threemagnetopoincare} shows that the curve $\omega_1(k_x)$ and the curve $\omega_1(k_x-k_{x1})+\omega_1(k_{x1})$ (the second term in the phase matching condition) can intersect so the surface $\omega_1(\mathbf{k})$ and the surface $\omega_1(\mathbf{k}-\mathbf{k_1})+\omega_1(\mathbf{k_1})$ can intersect along a curve. The intersection points correspond to magneto-Poincare wave with frequency equal to the frequencies sum and the wave vector is equal to the wave vectors sum of two waves. Thus the phase matching condition is satisfied for three magneto-Poincare waves, so their nonlinear interaction is allowed.

\begin{figure}
\center{\includegraphics{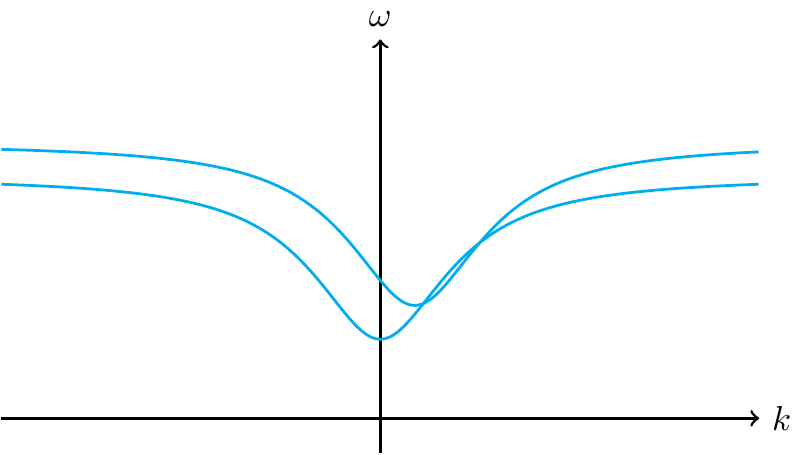}}
\caption{The intersection of the dispersion curves of magnetostrophic mode}
\label{pic.threemagnetostrophic}
\end{figure}

The phase matching conditions for three magnetostrophic waves are written as $\omega_2(\mathbf{k_1})+\omega_2(\mathbf{k_2})=\omega_2(\mathbf{k_1}+\mathbf{k_2})$. The subscript $2$ corresponds to the magnetostrophic wave. To understand whether this condition is realized let us turn to the figure \ref{pic.threemagnetostrophic}. The bottom curve $\omega_2(k_x)$ describes the first term in the phase matching condition. The upper curve  $\omega_2(k_x-k_{x1})+\omega_2(k_{x1})$ describes the second term.  The intersection of curves causes the intersection of dispersion surfaces. Thus there are such wave vectors $\mathbf{k_1},\mathbf{k_2}$ that the phase matching condition is satisfied.

\begin{figure}
\center{\includegraphics[scale=1.0]{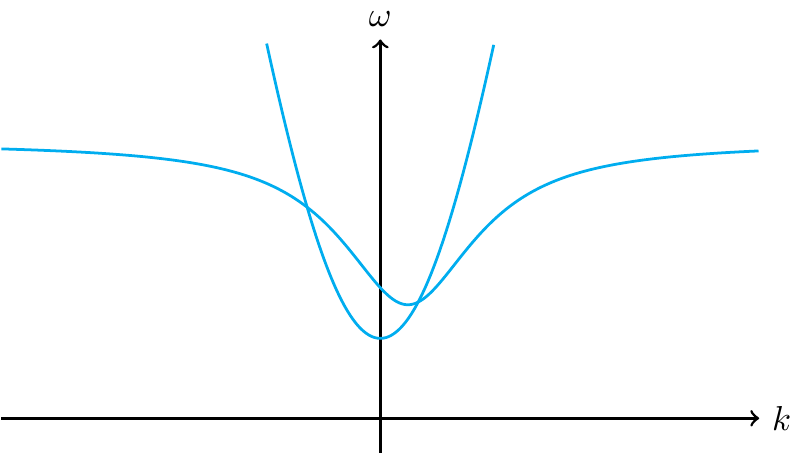}}
\caption{The intersection of dispersion curves of magnetostrophic and magneto-Poincare modes}
\label{pic.twomagnetostrophiconemagnetopoincare}
\end{figure}

\begin{figure}
\center{\includegraphics{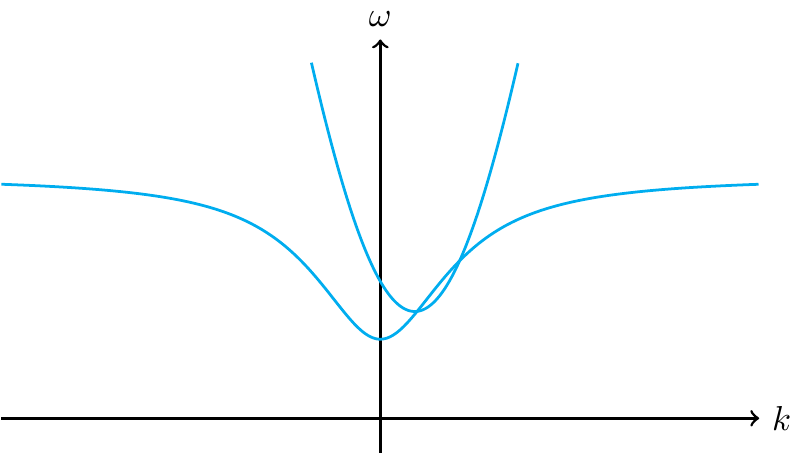}}
\caption{The intersection of dispersion curves of magneto-Poincare and magnetostrophic modes}
\label{pic.twomagnetopoincareonemagnetostrophic}
\end{figure}

In the case of interaction of one magnetostrophic wave and one magneto-Poincare wave magnetostrophic wave with following condition can turn: $\omega_2(\mathbf{k_1})+\omega_1(\mathbf{k_2})=\omega_2(\mathbf{k_1}+\mathbf{k_2})$ (figure \ref{pic.twomagnetostrophiconemagnetopoincare})or magneto-Poincare wave can turn and its phase matching condition is: $\omega_1(\mathbf{k_1})+\omega_2(\mathbf{k_2})=\omega_1(\mathbf{k_1}+\mathbf{k_2})$ (figure \ref{pic.twomagnetopoincareonemagnetostrophic}). In both cases the intersection of dispersion surfaces means that there are such $\mathbf{k_1}$ and $\mathbf{k_2}$ that will satisfy the phase matching conditions. Thus such three-wave interactions exist.

\section{Three-wave interactions in shallow water magnetohydrodynamics}
To study nonlinear effects we use multiscale asymptotic method \cite{ostrovsky2014asymptotic}, \cite{Pelinovsky}. The solution $\mathbf{u}=(h,v_x,v_y,B_x,B_y)$ of original equations (\ref{6}-\ref{10}) is represented in a form of a series in the small parameter $\epsilon$:
\begin{equation}\label{23}
\mathbf{u}=\epsilon\mathbf{u_0}+\epsilon^2\mathbf{u_1}+...
\end{equation}
Where $\mathbf{u_0}$ is linear solution of original system (\ref{6}-\ref{10}), $\mathbf{u_1}$ is a term describing quadratic nonlinearity effect and move from variables $(t,x,y)$ to fast $(T_0,X_0,Y_0)$ and slow $(T_1,X_1,Y_1)$ variables according to:
\begin{equation}\label{24}
\frac{\partial}{\partial t}=\frac{\partial}{\partial T_0}+\frac{\partial}{\partial T_1}+...
\end{equation}
\begin{equation}\label{25}
\frac{\partial}{\partial x}=\frac{\partial}{\partial X_0}+\frac{\partial}{\partial X_1}+...
\end{equation}
\begin{equation}\label{26}
\frac{\partial}{\partial y}=\frac{\partial}{\partial Y_0}+\frac{\partial}{\partial Y_1}+...
\end{equation}
Thus after introducing different scales the compatibility condition of equations system propotional to $\epsilon^2$ will be the equation on the slow varying amplitudes. This condition eliminates secular terms and defines $\mathbf{u_0}(T_1,X_1,Y_1)$. As a consequence we obtain following set of equations of waves interractions in shallow water MHD in the external vertical magnetic field. 

Considering three magneto-Poincare waves interaction with amplitudes $\alpha$, $\beta$ and $\gamma$  with the phase matching condition $\omega_1(\mathbf{k_1})+\omega_1(\mathbf{k_2})=\omega_1(\mathbf{k_1}+\mathbf{k_2})$ we obtain amplitude equations:
\begin{equation}\label{33}
a\frac{\partial \gamma}{\partial T}+b^1\frac{\partial \gamma}{\partial X}+b^2\frac{\partial \gamma}{\partial Y}=-i(k(\omega_1(\mathbf{k_1})+\omega_1(\mathbf{k_2}))+l^1(k_{x1}+k_{x2})+l^2(k_{y1}+k_{y2})+2m)\alpha \beta
\end{equation}
\begin{equation}\label{34}
a\frac{\partial \alpha}{\partial T}+b^1\frac{\partial \alpha}{\partial X}+b^2\frac{\partial \alpha}{\partial Y}=-i(k(\omega_1(\mathbf{k_3})-\omega_1(\mathbf{k_2}))+l^1(k_{x3}-k_{x2})+l^2(k_{y3}-k_{y2})+2m)\gamma \beta
\end{equation}
\begin{equation}\label{35}
a\frac{\partial \beta}{\partial T}+b^1\frac{\partial \beta}{\partial X}+b^2\frac{\partial \beta}{\partial Y}=-i(k(\omega_1(\mathbf{k_3})-\omega_1(\mathbf{k_1}))+l^1(k_{x3}-k_{x1})+l^2(k_{y3}-k_{y1})+2m)\gamma \alpha
\end{equation}
In these equations the coefficients $a$, $b^1$, $b^2$, $k$, $l^1$, $l^2$, $m$ are constant. They are determined by the initial conditions ($f$, $H$, $g$, $B_z$).

When considering the case when three magnetostrophic waves satisfy the phase matching condition $\omega_2(\mathbf{k_1})+\omega_2(\mathbf{k_2})=\omega_2(\mathbf{k_1}+\mathbf{k_2})$ we obtain following system describing slow amplitudes of three magnetostrophic waves:
\begin{equation}\label{39}
a\frac{\partial \chi}{\partial T}+b^1\frac{\partial \chi}{\partial X}+b^2\frac{\partial \chi}{\partial Y}=-i(k(\omega_2(\mathbf{k_1})+\omega_2(\mathbf{k_2}))+l^1(k_{x1}+k_{x2})+l^2(k_{y1}+k_{y2})+2m)\phi \psi
\end{equation}
\begin{equation}\label{40}
a\frac{\partial \phi}{\partial T}+b^1\frac{\partial \phi}{\partial X}+b^2\frac{\partial \phi}{\partial Y}=-i(k(\omega_2(\mathbf{k_3})-\omega_2(\mathbf{k_2}))+l^1(k_{x3}-k_{x2})+l^2(k_{y3}-k_{y2})+2m)\chi \psi
\end{equation}
\begin{equation}\label{41}
a\frac{\partial \psi}{\partial T}+b^1\frac{\partial \psi}{\partial X}+b^2\frac{\partial \psi}{\partial Y}=-i(k(\omega_2(\mathbf{k_3})-\omega_2(\mathbf{k_1}))+l^1(k_{x3}-k_{x1})+l^2(k_{y3}-k_{y1})+2m)\chi \phi
\end{equation}
In case when two magneto-Poincare waves and one magnetostrophic wave satisfy the phase matching condition $\omega_1(\mathbf{k_1})+\omega_2(\mathbf{k_2})=\omega_1(\mathbf{k_1}+\mathbf{k_2})$ we have:
\begin{multline}\label{42}
a\frac{\partial \beta}{\partial T}+b^1\frac{\partial \beta}{\partial X}+b^2\frac{\partial \beta}{\partial Y}= \\ =-i(k(\omega_2(\mathbf{k_1})+\omega_1(\mathbf{k_2}))+l^1(k_{x1}+k_{x2})+l^2(k_{y1}+k_{y2})+2m)\alpha \phi
\end{multline}
\begin{multline}\label{43}
a\frac{\partial \phi}{\partial T}+b^1\frac{\partial \phi}{\partial X}+b^2\frac{\partial \phi}{\partial Y}= \\ =-i(k(\omega_1(\mathbf{k_3})-\omega_1(\mathbf{k_2}))+l^1(k_{x3}-k_{x2})+l^2(k_{y3}-k_{y2})+2m)\beta \alpha
\end{multline}
\begin{multline}\label{44}
a\frac{\partial \alpha}{\partial T}+b^1\frac{\partial \alpha}{\partial X}+b^2\frac{\partial \alpha}{\partial Y}= \\ =-i(k(\omega_1(\mathbf{k_3})-\omega_2(\mathbf{k_1}))+l^1(k_{x3}-k_{x1})+l^2(k_{y3}-k_{y1})+2m)\beta \phi
\end{multline}
When two magnetostrophic waves and one magneto-Poincare wave interact $\omega_2(\mathbf{k_1})+\omega_1(\mathbf{k_2})=\omega_2(\mathbf{k_1}+\mathbf{k_2})$ we obtain:
\begin{multline}\label{81}
a\frac{\partial \psi}{\partial T}+b^1\frac{\partial \psi}{\partial X}+b^2\frac{\partial \psi}{\partial Y}= \\ =-i(k(\omega_1(\mathbf{k_1})+\omega_2(\mathbf{k_2}))+l^1(k_{x1}+k_{x2})+l^2(k_{y1}+k_{y2})+2m)\alpha \phi
\end{multline}
\begin{multline}\label{82}
a\frac{\partial \alpha}{\partial T}+b^1\frac{\partial \alpha}{\partial X}+b^2\frac{\partial \alpha}{\partial Y}= \\ =-i(k(\omega_2(\mathbf{k_3})-\omega_2(\mathbf{k_2}))+l^1(k_{x3}-k_{x2})+l^2(k_{y3}-k_{y2})+2m)\psi \phi
\end{multline}
\begin{multline}\label{83}
a\frac{\partial \phi}{\partial T}+b^1\frac{\partial \phi}{\partial X}+b^2\frac{\partial \phi}{\partial Y}= \\ =-i(k(\omega_2(\mathbf{k_3})-\omega_1(\mathbf{k_1}))+l^1(k_{x3}-k_{x1})+l^2(k_{y3}-k_{y1})+2m)\psi \alpha
\end{multline}

\section{Parametric instabilities of magneto-Poincare and magnetostrophic waves}

Here we use the obtained nonlinear three-wave interactions equations (three magneto-Poincare waves, three magnetostrophic waves, two magneto-Poincare and one magnetostropchic wave, two magnetostrophic waves and one magneto-Poincare wave) for the qualitative analysis of wave transformation fenomena in shallow water magnetohydrodynamics of astrophysical plasma in weakly nonlimear approximation \cite{falkovich2011fluid}.

\subsection{Decay instabilities}

In first case three magneto-Poincare waves satisfy the phase matching condition $\omega_1(\mathbf{k_1})+\omega_1(\mathbf{k_2})=\omega_1(\mathbf{k_1}+\mathbf{k_2})$ and thus interact. Considering the initial conditions in the equations (\ref{33}-\ref{35}) when the amplitude of one wave is much higher than the amplitudes of other waves $\gamma=\gamma_0>>\alpha,\beta$ we obtain the linear differential equations with the argument $(\alpha,\beta)$. The obtained solution displays amplitude growth with the following increment:
\begin{equation}\label{raspad1}
\Gamma=\frac{|f^*| \gamma_0}{a}
\end{equation}
where
\begin{equation}
f^*=-i(k(\omega_1(\mathbf{k_3})-\omega_1(\frac{\mathbf{k_3}}{2}))+l^1(\frac{k_{x3}}{2})+l^2(\frac{k_{y3}}{2})+2m)
\end{equation}
Thus, we showed that one magneto-Poincare wave with a frequency $\omega_3$ and the wave vector $\mathbf{k_3}$ decays into two magneto-Poncare waves with frequences $\frac{\omega_3}{2}$ and wave vectors $\frac{\mathbf{k_3}}{2}$ in MHD of astrophysical plasma in external magnetic field.

The second case is three magnetostrophic waves interaction which satisfy the phase matching condition $\omega_2(\mathbf{k_1})+\omega_2(\mathbf{k_2})=\omega_2(\mathbf{k_1}+\mathbf{k_2})$. Considering the initial conditions in the equations (\ref{39}-\ref{41}) when the amplitude of one wave is much higher than the amplitudes of other waves $\chi=\chi_0>>\phi,\psi$ we obtain amplitude growth with the following increment:
\begin{equation}\label{raspad2}
\Gamma=\frac{f^*\chi_0}{a}
\end{equation}
where
\begin{equation}
f^*=-i(k(\omega_2(\mathbf{k_3})-\omega_2(\frac{\mathbf{k_3}}{2}))+l^1(\frac{k_{x3}}{2})+l^2(\frac{k_{y3}}{2})+2m)
\end{equation}
Thus, magnetostrophic wave with frequency $\omega_3$ and wave vector $\mathbf{k_3}$ decays into two magnetostrophic waves with frequences $\frac{\omega_3}{2}$ and wave vectors $\frac{\mathbf{k_3}}{2}$ in shallow water MHD of astrophysical plasma.

Now let us consider two magneto-Poincare and one magnetostrophic wave interaction which satisfy the phase matching condition $\omega_1(\mathbf{k_1})+\omega_2(\mathbf{k_2})=\omega_1(\mathbf{k_1}+\mathbf{k_2})$. In this case the initial conditions in the equations (\ref{42}-\ref{44}) are: the amplitude of one magneto-Poincare wave is much higher than the amplitudes of other waves $\beta=\beta_0>>\phi,\alpha$ we obtain the linear system of differential equations with the argument $(\phi,\alpha)$. The obtained solution displays amplitude growth with the following increment: 
\begin{equation}\label{raspad3}
\Gamma= \frac{\sqrt{|f_2^*| |f_3^*|} \beta_0}{a}
\end{equation}
where
\begin{equation}
f_2^*=-i(k(\omega_1(\mathbf{k_3})-\omega_1(\mathbf{k_3}-\mathbf{k_1^0}))+l^1k_{x1}^0)+l^2k_{y1}^0)+2m)
\end{equation}
\begin{equation}
f_3^*=-i(k(\omega_1(\mathbf{k_3})-\omega_2(\mathbf{k_1^0}))+l^1(k_{x3}-k_{x1}^0)+l^2(k_{y3}-k_{y1}^0)+2m)
\end{equation}
Thus, magneto-Poincare wave with a frequency $\omega_3$ and wave vector $\mathbf{k_3}$ decays into magnetostrophic wave with a frequency $\omega_1^0$ and wave vector $\mathbf{k_1^0}$ and magneto-Poicare wave with frequency $\omega_3-\omega_1^0$ and wave vector $\mathbf{k_3}-\mathbf{k_1^0}$ where $\mathbf{k_1^0}$ is determined from the constants in free surface MHD flows in shallow water approximation.

Finally, two magnetostrophic waves and one magneto-Poincare wave interact and satisfy the phase matching condition $\omega_2(\mathbf{k_1})+\omega_1(\mathbf{k_2})=\omega_2(\mathbf{k_1}+\mathbf{k_2})$. Considering the initial conditions in the equations (\ref{81}-\ref{83}) when the amplitude of one magnetostrophic wave is much higher than the amplitudes of other waves $\psi=\psi_0>>\phi,\alpha$ we obtain similarly amplitude growth with the following increment:
\begin{equation}\label{raspad4}
\Gamma=\frac{\sqrt{|f_2^*| |f_3^*|} \psi_0}{a}
\end{equation}
where
\begin{equation}
f_2^*=-i(k(\omega_2(\mathbf{k_3})-\omega_2(\mathbf{k_3}-\mathbf{k_1^0}))+l^1(k_{x1}^0)+l^2(k_{y1}^0)+2m)
\end{equation}
\begin{equation}
f_3^*=-i(k(\omega_2(\mathbf{k_3})-\omega_1(\mathbf{k_1^0}))+l^1(k_{x3}-k_{x1}^0)+l^2(k_{y3}-k_{y1}^0)+2m)
\end{equation}
Thus, we showed that magnetostrophic wave with a frequency $\omega_3$ and wave vector $\mathbf{k_3}$ decays into magneto-Poincare wave with frequency $\omega_1^0$ and wave vector $\mathbf{k_1^0}$ and magnetostrophic wave with frequency $\omega_3-\omega_1^0$ and wave vector $\mathbf{k_3}-\mathbf{k_1^0}$ where $\mathbf{k_1^0}$ is determined from the constants in shallow water MHD astrophysical plasma in the external magnetic field.

\subsection{Parametric amplification}
Another type of instabilities that is described by nonlinear equations is parametric amplification. It represent the amplification of third wave as a result of nonlinear interaction of two initially propagating waves.

Let us again consider three magneto-Poincare waves interaction which sastisfy the phase matching condition $\omega_1(\mathbf{k_1})+\omega_1(\mathbf{k_2})=\omega_1(\mathbf{k_1}+\mathbf{k_2})$. The initial condition in equations (\ref{33}-\ref{35}) is that the amplitude of one wave is much smaller than the amplitudes of two other waves $\gamma<<\alpha_0,\beta_0$. This condition causes the linear differential equation with the argument $\gamma$. The obtained solution displays amplitude growth with the following increment:
\begin{equation}\label{usilenie1}
\Gamma=\frac{|f_1| \alpha_0 \beta_0}{a}
\end{equation}
where
\begin{equation}
f_1=-i(k(\omega_1(\mathbf{k_1})+\omega_1(\mathbf{k_2}))+l^1(k_{x1}+k_{x2})+l^2(k_{y1}+k_{y2})+2m)
\end{equation}
Thus, we showed that amplifying magneto-Poincare wave has  frequency  $\omega_3=\omega_1+\omega_2$ and wave vector $\mathbf{k_3}=\mathbf{k_1}+\mathbf{k_2}$ in MHD of astrophysical plasma in external magnetic field.

When considering three magnetostrophic wave interaction which satisfy the phase matching condition $\omega_2(\mathbf{k_1})+\omega_2(\mathbf{k_2})=\omega_2(\mathbf{k_1}+\mathbf{k_2})$ the initial condition in equations (\ref{39}-\ref{41}) is that the amplitude of one wave is much smaller than the amplitudes of two other waves $\chi<<\psi_0,\phi_0$. This condition causes the linear differential equation with the argument $\chi$. The obtained solution displays amplitude growth with the following increment:
\begin{equation}\label{usilenie2}
\Gamma=\frac{|f_1| \psi_0 \phi_0}{a}
\end{equation}
where
\begin{equation}
f_1=-i(k(\omega_2(\mathbf{k_1})+\omega_2(\mathbf{k_2}))+l^1(k_{x1}+k_{x2})+l^2(k_{y1}+k_{y2})+2m)
\end{equation}
Thus, amplifying magnetostrophic wave has frequency $\omega_3=\omega_1+\omega_2$ and wave vector $\mathbf{k_3}=\mathbf{k_1}+\mathbf{k_2}$ in shallow water MHD of astrophysical plasma in the external magnetic field.

If we consider two magneto-Poincare waves and one magnetostrophic wave interaction which satisfy the phase matching condition $\omega_1(\mathbf{k_1})+\omega_2(\mathbf{k_2})=\omega_1(\mathbf{k_1}+\mathbf{k_2})$. The initial condition in equations (\ref{42}-\ref{44}) is that the amplitude of one wave is much smaller than the amplitudes of two other waves $\beta<<\alpha_0,\phi_0$. This condition causes the linear differential equation with the argument $\beta$. The obtained solution displays amplitude growth with the following increment:
\begin{equation}\label{usilenie3}
\Gamma=\frac{|f_1| \alpha_0 \phi_0}{a}
\end{equation}
where
\begin{equation}
f_1=-i(k(\omega_2(\mathbf{k_1})+\omega_1(\mathbf{k_2}))+l^1(k_{x1}+k_{x2})+l^2(k_{y1}+k_{y2})+2m)
\end{equation}
Thus, magneto-Poincare wave is the sum of magnetostrophic wave with frequency and wave vector $\omega_1$ and $\mathbf{k_1}$ and magneto-Poincare wave with frequency and wave vector $\omega_2$ and $\mathbf{k_2}$. The resulting wave frequency is $\omega_3=\omega_1+\omega_2$, its wave vector is $\mathbf{k_3}=\mathbf{k_1}+\mathbf{k_2}$ in free surface flows in shallow water MHD of astrophysical plasma.

The last case of the three-wave interactions is two magnetostrophic waves and one magneto-Poicare wave interaction which satisfy the phase matching condition $\omega_2(\mathbf{k_1})+\omega_1(\mathbf{k_2})=\omega_2(\mathbf{k_1}+\mathbf{k_2})$. The initial condition in equations (\ref{81}-\ref{83}) is that the amplitude of one wave is much smaller than the amplitudes of two other waves $\psi<<\alpha_0,\phi_0$. This condition causes the linear differential equation with the argument $\psi$. The obtained solution displays amplitude growth with the following increment:
\begin{equation}\label{usilenie4}
\Gamma=\frac{|f_1| \phi_0 \alpha_0}{a}
\end{equation}
where
\begin{equation}
f_1=-i(k(\omega_1(\mathbf{k_1})+\omega_2(\mathbf{k_2}))+l^1(k_{x1}+k_{x2})+l^2(k_{y1}+k_{y2})+2m)
\end{equation}
Therefore, amplified magnetostrophic wave is the sum of magneto-Poincare wave with frequency $\omega_1$ and wave vector $\mathbf{k_1}$ and magnetostrophic wave with frequency $\omega_2$ and wave vector $\mathbf{k_2}$. The resulting wave frequency is $\omega_3=\omega_1+\omega_2$ and its wave vector is $\mathbf{k_3}=\mathbf{k_1}+\mathbf{k_2}$.

\section{Conclusion}
The weakly nonlinear theory of wave interactions in large scale flows of rotating plasma layer with a free surface in gravity field on a flat surface in vertical magnetic field is developed. We use MHD shallow water approximation to obtain the wave packets interaction equations of magneto-Poincare and magnetostrophic waves. By analyzing the dispersion surfaces of investigated waves it is found that such condition can be satisfied for following configurations: three magneto-Poncare waves (figure \ref{pic.threemagnetopoincare}), three magnetostrophic waves (figure \ref{pic.threemagnetostrophic}), and also two magnetostrophic waves and one magneto-Poicare wave (figure \ref{pic.twomagnetostrophiconemagnetopoincare}) and two magneto-Poincare waves and one magnetostrophic wave (figure \ref{pic.twomagnetopoincareonemagnetostrophic}). We obtain weakly nonlinear solution by asymptotic multiscale method. For every obtained type of interactions the nonlinear equations sytems are found. It is shown that in weakly nonlinear approximation there are decay instabilities and parametric amplifications. It's found that there are following types of decay instabilities: magneto-Poincare wave decays into two magneto-Poincare waves with the increment (\ref{raspad1}), magnetostrophic wave decays into two magnetostrophic waves with the increment (\ref{raspad2}), magneto-Poincare wave decays into one magneto-Poincare wave and one magnetostrophic wave with the increment (\ref{raspad3}), magnetostrophic wave decays into one magnetostrophic wave and one magneto-Poincare wave with the increment (\ref{raspad4}). Also there are four types of parametric amplifications: parametric amplification of the magneto-Poincare waves with instability growth rate (\ref{usilenie1}), parametric amplification of magnetostrophic waves with instability growth rate (\ref{usilenie2}), and also parametric amplification of magneto-Poincare wave in magneto-Poincare wave field with instability growth rate (\ref{usilenie3}) and amplification of magnetostrophic wave in magneto-Poincare wave field with instability growth rate (\ref{usilenie4}). The work is supported by the program 9 of RAS presidium "Experimental and theoretical studies of solar system objects and star planetary systems" and RFFI grant 14-29-06065. The authors are grateful to S.Y. Dobrohotov for useful discussions about asymptotic multiscale methods.

\nocite{*}
\end{document}